\documentclass[10pt,aps,prl,reprint,twocolumn,amsmath,superscriptaddress,preprintnumbers,floatfix,nofootinbib,showpacs]{revtex4-1}

\usepackage{siunitx}
\usepackage{tikz}
\usepackage{tikz-3dplot}
\usepackage{graphicx}
\usepackage{xcolor}
\usepackage[colorlinks=false, pdfborder={0 0 0}]{hyperref}
\usepackage{mathtools}
\usepackage{amssymb}
\usepackage[utf8]{inputenc}
\usepackage[T1]{fontenc}
\usepackage{natbib}

\usepackage{ulem}

\begin{document}
\title{Phase locking the spin precession in a storage ring}

\author{N.~Hempelmann}
\affiliation{III. Physikalisches Institut B, RWTH Aachen University, 52056 Aachen, Germany}
\affiliation{Institut f\"ur Kernphysik, Forschungszentrum J\"ulich, 52425 J\"ulich, Germany}
\author{V.~Hejny}
\affiliation{Institut f\"ur Kernphysik, Forschungszentrum J\"ulich, 52425 J\"ulich, Germany}
\author{J.~Pretz}
\affiliation{III. Physikalisches Institut B, RWTH Aachen University, 52056 Aachen, Germany}
\affiliation{Institut f\"ur Kernphysik, Forschungszentrum J\"ulich, 52425 J\"ulich, Germany}
\affiliation{JARA--FAME (Forces and Matter Experiments), Forschungszentrum J\"ulich and RWTH Aachen University, Germany}
\author{E.~Stephenson}
\affiliation{Indiana University Center for Spacetime Symmetries, Bloomington,  Indiana 47405, USA}
\author{W.~Augustyniak}
\affiliation{Department of Nuclear Physics, National Centre for Nuclear Research, 00681 Warsaw, Poland}
\author{Z.~Bagdasarian}
\affiliation{High Energy Physics Institute, Tbilisi State University, 0186 Tbilisi, Georgia}
\affiliation{Institut f\"ur Kernphysik, Forschungszentrum J\"ulich, 52425 J\"ulich, Germany}
\author{M.~Bai}
\affiliation{Institut f\"ur Kernphysik, Forschungszentrum J\"ulich, 52425 J\"ulich, Germany}
\affiliation{JARA--FAME (Forces and Matter Experiments), Forschungszentrum J\"ulich and RWTH Aachen University, Germany}
\author{L.~Barion}
\affiliation{University of Ferrara and INFN, 44100 Ferrara, Italy}
\author{M.~Berz}
\affiliation{Department of Physics and Astronomy, Michigan State University,  East Lansing, Michigan 48824, USA}
\author{S.~Chekmenev}
\affiliation{III. Physikalisches Institut B, RWTH Aachen University, 52056 Aachen, Germany}
\author{G.~Ciullo}
\affiliation{University of Ferrara and INFN, 44100 Ferrara, Italy}
\author{S.~Dymov}
\affiliation{Institut f\"ur Kernphysik, Forschungszentrum J\"ulich, 52425 J\"ulich, Germany}
\affiliation{Laboratory of Nuclear Problems, Joint Institute for Nuclear Research, 141980 Dubna, Russia}
\author{F.-J.~Etzkorn}
\affiliation{Institut f\"ur Kernphysik, Forschungszentrum J\"ulich, 52425
J\"ulich, Germany}
\author{D. Eversmann}
\affiliation{III. Physikalisches Institut B, RWTH Aachen University, 52056 Aachen, Germany}
\author{M.~Gaisser}
\affiliation{III. Physikalisches Institut B, RWTH Aachen University, 52056 Aachen, Germany}
\affiliation{Center for Axion and Precision Physics Research, Institute for Basic Science (IBS), Daejeon 34141, Republic of Korea}
\author{R.~Gebel}
\affiliation{Institut f\"ur Kernphysik, Forschungszentrum J\"ulich, 52425 J\"ulich, Germany}
\author{K.~Grigoryev}
\affiliation{III. Physikalisches Institut B, RWTH Aachen University, 52056 Aachen, Germany}
\author{D.~Grzonka}
\affiliation{Institut f\"ur Kernphysik, Forschungszentrum J\"ulich, 52425 J\"ulich, Germany}
\author{G.~Guidoboni}
\affiliation{University of Ferrara and INFN, 44100 Ferrara, Italy}
\author{T.~Hanraths}
\affiliation{Institut f\"ur Kernphysik, Forschungszentrum J\"ulich, 52425 J\"ulich, Germany}
\author{D.~Heberling}
\affiliation{Institut f\"ur Hochfrequenztechnik, RWTH Aachen University, 52056 Aachen, Germany}
\affiliation{JARA--FAME (Forces and Matter Experiments), Forschungszentrum J\"ulich and RWTH Aachen University, Germany}
\author{J.~Hetzel}
\affiliation{Institut f\"ur Kernphysik, Forschungszentrum J\"ulich, 52425 J\"ulich, Germany}
\author{F.~Hinder}
\affiliation{III. Physikalisches Institut B, RWTH Aachen University, 52056 Aachen, Germany}
\affiliation{Institut f\"ur Kernphysik, Forschungszentrum J\"ulich, 52425 J\"ulich, Germany}
\author{A.~Kacharava}
\affiliation{Institut f\"ur Kernphysik, Forschungszentrum J\"ulich, 52425 J\"ulich, Germany}
\author{V.~Kamerdzhiev}
\affiliation{Institut f\"ur Kernphysik, Forschungszentrum J\"ulich, 52425 J\"ulich, Germany}
\author{I.~Keshelashvili}
\affiliation{Institut f\"ur Kernphysik, Forschungszentrum J\"ulich, 52425 J\"ulich, Germany}
\author{I.~Koop}
\affiliation{Budker Institute of Nuclear Physics, 630090 Novosibirsk, Russia}
\author{A.~Kulikov}
\affiliation{Laboratory of Nuclear Problems, Joint Institute for Nuclear Research, 141980 Dubna, Russia}
\author{A.~Lehrach}
\affiliation{III. Physikalisches Institut B, RWTH Aachen University, 52056 Aachen, Germany}
\affiliation{Institut f\"ur Kernphysik, Forschungszentrum J\"ulich, 52425 J\"ulich, Germany}
\affiliation{JARA--FAME (Forces and Matter Experiments), Forschungszentrum J\"ulich and RWTH Aachen University, Germany}
\author{P.~Lenisa}
\affiliation{University of Ferrara and INFN, 44100 Ferrara, Italy}
\author{N.~Lomidze}
\affiliation{High Energy Physics Institute, Tbilisi State University, 0186 Tbilisi, Georgia}
\author{B.~Lorentz}
\affiliation{Institut f\"ur Kernphysik, Forschungszentrum J\"ulich, 52425 J\"ulich, Germany}
\author{P.~Maanen}
\affiliation{III. Physikalisches Institut B, RWTH Aachen University, 52056 Aachen, Germany}
\author{G.~Macharashvili}
\affiliation{High Energy Physics Institute, Tbilisi State University, 0186 Tbilisi, Georgia}
\affiliation{Laboratory of Nuclear Problems, Joint Institute for Nuclear Research, 141980 Dubna, Russia}
\author{A.~Magiera}
\affiliation{Institute of Physics, Jagiellonian University, 30348 Cracow, Poland}
\author{D.~Mchedlishvili}
\affiliation{High Energy Physics Institute, Tbilisi State University, 0186 Tbilisi, Georgia}
\affiliation{Institut f\"ur Kernphysik, Forschungszentrum J\"ulich, 52425 J\"ulich, Germany}
\author{S.~Mey}
\affiliation{III. Physikalisches Institut B, RWTH Aachen University, 52056 Aachen, Germany}
\affiliation{Institut f\"ur Kernphysik, Forschungszentrum J\"ulich, 52425 J\"ulich, Germany}
\author{F.~M\"uller}
\affiliation{III. Physikalisches Institut B, RWTH Aachen University, 52056 Aachen, Germany}
\affiliation{Institut f\"ur Kernphysik, Forschungszentrum J\"ulich, 52425 J\"ulich, Germany}
\author{A.~Nass}
\affiliation{Institut f\"ur Kernphysik, Forschungszentrum J\"ulich, 52425 J\"ulich, Germany}
\author{N.N. Nikolaev}
\affiliation{L.D. Landau Institute for Theoretical Physics, 142432 Chernogolovka, Russia}
\affiliation{Moscow Institute for Physics and Technology, 141700 Dolgoprudny, Russia}
\author{A.~Pesce}
\affiliation{University of Ferrara and INFN, 44100 Ferrara, Italy}
\author{D.~Prasuhn}
\affiliation{Institut f\"ur Kernphysik, Forschungszentrum J\"ulich, 52425 J\"ulich, Germany}
\author{F.~Rathmann}
\affiliation{Institut f\"ur Kernphysik, Forschungszentrum J\"ulich, 52425
  J\"ulich, Germany}
\author{M.~Rosenthal}
\affiliation{III. Physikalisches Institut B, RWTH Aachen University, 52056 Aachen, Germany}
\affiliation{Institut f\"ur Kernphysik, Forschungszentrum J\"ulich, 52425 J\"ulich, Germany}
\author{A.~Saleev}
\affiliation{Institut f\"ur Kernphysik, Forschungszentrum J\"ulich, 52425 J\"ulich, Germany}
\affiliation{Samara National Research University, 443086 Samara, Russia}
\author{V.~Schmidt}
\affiliation{III. Physikalisches Institut B, RWTH Aachen University, 52056 Aachen, Germany}
\affiliation{Institut f\"ur Kernphysik, Forschungszentrum J\"ulich, 52425 J\"ulich, Germany}
\author{Y.~Semertzidis}
\affiliation{Center for Axion and Precision Physics Research, Institute for Basic Science (IBS), Daejeon 34141, Republic of Korea}

\affiliation{Department of Physics, Korea Advanced Institute of Science and Technology (KAIST), Daejeon 34141, Republic of Korea}

%
%
\author{V.~Shmakova}
\affiliation{Laboratory of Nuclear Problems, Joint Institute for Nuclear Research, 141980 Dubna, Russia}
\author{A.~Silenko}
\affiliation{Research Institute for Nuclear Problems, Belarusian State University, 220030 Minsk, Belarus}
\affiliation{Bogoliubov Laboratory of Theoretical Physics, Joint Institute for Nuclear Research, 141980 Dubna, Russia}
\author{J.~Slim}
\affiliation{Institut f\"ur Hochfrequenztechnik, RWTH Aachen University, 52056 Aachen, Germany}
\author{H.~Soltner}
\affiliation{Zentralinstitut f\"ur Engineering, Elektronik und Analytik (ZEA-1), Forschungszentrum J\"ulich, 52425 J\"ulich, Germany}
\author{A.~Stahl}
\affiliation{III. Physikalisches Institut B, RWTH Aachen University, 52056 Aachen, Germany}%
\affiliation{JARA--FAME (Forces and Matter Experiments), Forschungszentrum J\"ulich and RWTH Aachen University, Germany}
\author{R.~Stassen}
\affiliation{Institut f\"ur Kernphysik, Forschungszentrum J\"ulich, 52425 J\"ulich, Germany}
\author{H.~Stockhorst}
\affiliation{Institut f\"ur Kernphysik, Forschungszentrum J\"ulich, 52425 J\"ulich, Germany}
\author{H.~Str\"oher}
\affiliation{Institut f\"ur Kernphysik, Forschungszentrum J\"ulich, 52425 J\"ulich, Germany}
\affiliation{JARA--FAME (Forces and Matter Experiments), Forschungszentrum J\"ulich and RWTH Aachen University, Germany}
\author{M.~Tabidze}
\affiliation{High Energy Physics Institute, Tbilisi State University, 0186 Tbilisi, Georgia}
\author{G.~Tagliente}
\affiliation{INFN, 70125 Bari, Italy}
\author{R.~Talman}
\affiliation{Cornell University, Ithaca,  New York 14850, USA}
\author{P.~Th\"orngren Engblom}
\affiliation{Department of Physics, KTH Royal Institute of Technology, SE-10691 Stockholm, Sweden}
%
\author{F.~Trinkel}
\affiliation{III. Physikalisches Institut B, RWTH Aachen University, 52056 Aachen, Germany}
\affiliation{Institut f\"ur Kernphysik, Forschungszentrum J\"ulich, 52425 J\"ulich, Germany}
\author{Yu.~Uzikov}
\affiliation{Laboratory of Nuclear Problems, Joint Institute for Nuclear Research, 141980 Dubna, Russia}
\author{Yu.~Valdau}
\affiliation{Helmholtz-Institut f\"ur Strahlen- und Kernphysik, Universit\"at Bonn, 53115 Bonn, Germany}
\affiliation{Petersburg Nuclear Physics Institute, 188300 Gatchina, Russia}
\author{E.~Valetov}
\affiliation{Department of Physics and Astronomy, Michigan State University,  East Lansing, Michigan 48824, USA}
\author{A.~Vassiliev}
\affiliation{Petersburg Nuclear Physics Institute, 188300 Gatchina, Russia}
\author{C.~Weidemann}
\affiliation{University of Ferrara and INFN, 44100 Ferrara, Italy}
\author{A.~Wro\'{n}ska}
\affiliation{Institute of Physics, Jagiellonian University, 30348 Cracow, Poland}
\author{P.~W\"ustner}
\affiliation{Zentralinstitut f\"ur Engineering, Elektronik und Analytik (ZEA-2), Forschungszentrum J\"ulich, 52425 J\"ulich, Germany}
\author{P.~Zupra\'nski}
\affiliation{Department of Nuclear Physics, National Centre for Nuclear Research, 00681 Warsaw, Poland}
\author{M.~$\dot {\mathrm Z}$urek}
\affiliation{Institut f\"ur Kernphysik, Forschungszentrum J\"ulich, 52425 J\"ulich, Germany}
\collaboration{JEDI collaboration}


\begin{abstract}
This letter reports the successful use of feedback from a spin polarization
measurement to the revolution frequency of a 0.97~GeV/$c$ bunched and polarized
deuteron beam in the Cooler Synchrotron (COSY) storage ring in order to control both the precession
rate ($\approx 121$~kHz) and  the phase of the horizontal polarization component. Real
time synchronization with a radio frequency (rf) solenoid 
made possible the rotation of the polarization out of the horizontal plane,
yielding a demonstration of the feedback method to manipulate
the polarization. In particular, the rotation rate shows a sinusoidal function
of the horizontal polarization phase (relative to the rf solenoid), which was controlled to
within a one standard deviation range of $\sigma = 0.21$~rad. The minimum
possible adjustment was 3.7~mHz out of a revolution frequency of 753 kHz, which changes the precession rate by 26~mrad/s. 
Such a capability meets a requirement for the use of storage rings to look for
an intrinsic electric dipole moment of charged particles.
\end{abstract}

\maketitle
Over the last few decades, the importance of polarized beams has been demonstrated in numerous experiments at colliders as well as at storage rings such as the Cooler Synchroton COSY. Recently the emphasis of such experiments has started to shift from nuclear and hadron physics to precision measurements. This letter describes the successful effort to use measurements of the polarization of a storage ring deuteron beam in a feedback loop 
to control in real time the rate of precession in the horizontal (ring) plane
and the phase of that rotating polarization relative 
to an external reference. This work was carried out in the context of the
effort to search for an intrinsic electric dipole moment (EDM) 
of charged particles circulating in the ring~\cite{jedi}. 
Applications in other fields include the manipulation
of the polarisation vector, where
regular spin rotators are not practical.

An EDM aligned along the
particle spin axis is CP-violating and any observation of such a moment 
would be a signal of new physical processes possibly related to the
matter-antimatter asymmetry of the universe~\cite{Engel201321,Canetti:2012zc}.
An EDM may be observed by measuring the rate of spin precession in an external
electric field. 
For charged particles, confinement in a storage ring exposes them to the
particle rest frame electric field that closes the storage ring orbit. 
To work best for small EDMs, the beam polarization should begin parallel to
the particle velocity (and perpendicular to the radial electric field) 
and precess, due to the EDM, into the vertical direction where that polarization
component may be measured in a scattering experiment. 
Since the precession due to the magnetic moment of a particle is
different in electric and magnetic fields (compared to the amount by which the
particle path is bent), 
some combination of $E$ and $B$ bending fields may be found that allows the reference
particle to travel in a “frozen-spin” configuration in which the polarization
and the velocity remain aligned. But maintaining this alignment over a
thousand seconds~\cite{Guidoboni:2016bdn} for a statistically significant measurement
requires control of the rotation rate at the level of parts per billion; hence
real time feedback from a continuously running polarization measurement
(available at high precision~\cite{Eversmann:2015jnk}) is mandatory for any such EDM
experiment.

The frozen-spin configuration is not the only possibility to observe an EDM.
It has been proposed that 
precession of the EDM may be achieved through the use of a radio frequency (rf) Wien filter
with a vertical magnetic field~\cite{PhysRevSTAB.16.114001,Slim:2016pim} 
operating at a harmonic frequency of the horizontal spin precession in the
plane of a purely magnetic storage ring.
In this case both the precession rate and phase relative to the Wien filter rf
signal must be controlled. 
Here, we report on the first successful installation and operation of
a feedback system fulfilling these requirements.

The measurements presented here were performed at the Cooler Synchrotron 
COSY at the Forschungszentrum J\"ulich~\cite{Maier:1997zj}.
A vector polarized deuteron beam was injected at time $t=0$ and  
accelerated to a momentum of 970~MeV/$c$. The beam intensity was 
approximately $10^{9}$ deuterons per fill. The vector polarization
was perpendicular to the ring plane and was alternated 
from cycle to cycle from upward to downward direction. The corresponding values for the polarization were 
$p^{\textrm{up}} = 0.30\pm0.03$ and $p^{\textrm{down}}=-0.46\pm0.03$. 
After acceleration the beam was 
electron-cooled for $\SI{74}{s}$ to reduce the beam emittance. 
A radio frequency cavity to bunch the beam was used throughout the $\SI{200}{s}$ long 
cycle. 
Starting at $t=\SI{80}{s}$ the beam was slowly extracted onto an 
internal carbon target by applying a white noise electric field.
Elastically scattered deuterons were detected in scintillation detectors 
consisting of rings and bars around the beam pipe. The detector covers a 
range from $9$ to $13$ degrees in polar angle and is
segmented in four regions in the azimuthal angle (up, down, left and 
right).
The elasticity of the event was guaranteed by stopping the deuterons in the
outer scintillator ring and measuring their energy deposition
\cite{PhysRevSTAB.17.052803}.
The left-right asymmetry gives access to the vertical polarization,
the up-down asymmetry is sensitive to the polarization perpendicular to the
momentum vector in the horizontal plane of the ring.
At $t=\SI{85}{s}$ an rf solenoid was used to rotate the polarization 
from the initially vertical direction into the horizontal 
plane. 

Once the polarization is rotated out of the vertical direction, its horizontal
component starts to precess around the vertical axis. Defining the spin motion
in the rest frame of the particle,
the precession frequency is 
\begin{equation}
f_\mathrm{s} = \nu_s f_\mathrm{cosy} = \gamma G f_\mathrm{cosy} \, ,
\end{equation}
where $\nu_s = \gamma G \approx -0.16$ is the spin tune, $\gamma$ the Lorentz
factor, $G$ the gyromagnetic anomaly
and $f_\mathrm{cosy}$ the revolution frequency. 
Note that the sign of $f_\mathrm{s}$ indicates the direction of the spin rotation.
In order to rotate the polarization vector effectively, the rf solenoid has to run at a resonance frequency, which is given as
\begin{equation}
f_\mathrm{rf} = ( k + \nu_s ) f_\mathrm{cosy}, k \in \mathbb{Z} \, .
\label{eq:resonance_condition}
\end{equation}
For this experiment the solenoid was operated at $|f_\mathrm{rf}| \approx
\SI{873}{kHz}$ ($k=-1$). Table~\ref{tab:parameters} summarizes the most important parameters of the experiment.

In the further course of the cycle the solenoid was also used to restore the vertical 
polarization, thus providing a test for the feedback system. As the solenoid field
only acts on the polarization components perpendicular to the
solenoid field, which is aligned along the beam direction, its effective strength
is largest if the maximum of the rf solenoid field coincides in time with the 
horizontal polarization being radial and vanishes if both are parallel. 
The rate of build-up of vertical polarization 
is thus proportional to $\sin\phi$, 
where $\phi$ is the relative phase between the 
rf signal and the spin precession at the location of the rf solenoid.
$\phi=0$ corresponds to the case where the polarization vector 
is parallel to the magnetic field of the solenoid when it is 
at its maximum value. 
The slope of the polarization build-up as a function of $\phi$ is thus a measure how well the relative
phase between the two is under control.
For this, the solenoid was again switched on at $t=115\,$s at low field strength.

\begin{table}
\begin{center}
\begin{tabular}{|l|l|}
\hline\hline
deuteron momentum $p$       &  $\SI{0.970}{\,GeV\per c}$ \\
rel. velocity  $\beta$     & 0.459 \\
Lorentz factor $\gamma$ &   1.126 \\
slip factor $\eta$ & -0.58 \\
gyromagnetic anomaly $G$ & $\approx -0.143$ \\
revolution frequency $f_{\mathrm{cosy}}$ & 752543\, Hz\\
precession frequency $|f_{\mathrm{s}}|$ & 121173\, Hz\\
resonance frequency $|f_{\mathrm{rf}}|$ & 873716\, Hz\\
cycle length &  200 \,s  \\
\hline\hline
\end{tabular}
\caption{Parameters of the experiment.\label{tab:parameters}}
\end{center}
\end{table}

\begin{figure}
  \includegraphics[width=\columnwidth]{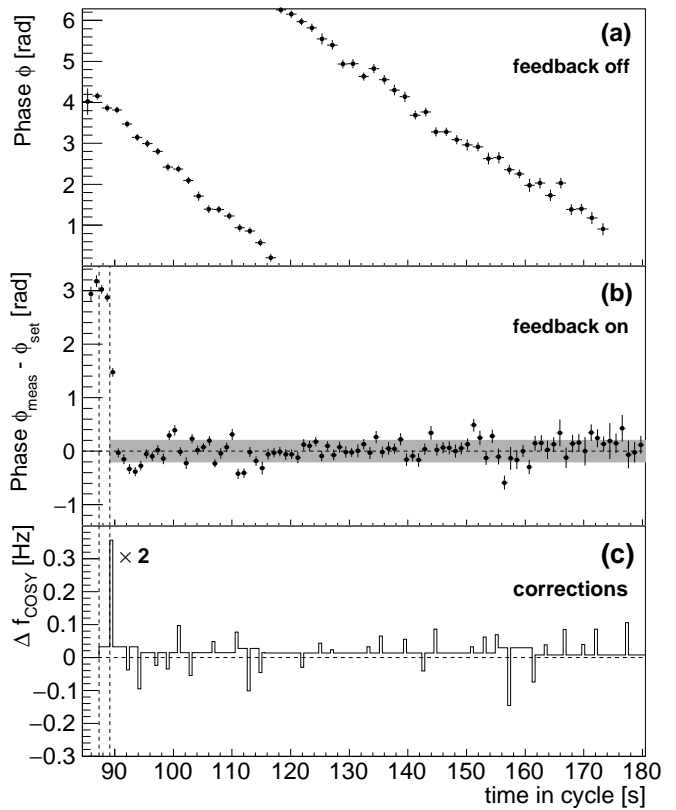}
  \caption{Phase between the measured polarization direction and the rf signal
           of the solenoid (a) without and (b) with an active feedback 
           system. The y-axis in (b) has been shifted to the predefined
           value. The gray band indicates the $\pm 1\sigma$ band for the 
           resulting phase distribution. In (c) the corresponding corrections to the revolution
           frequency $f_{\mathrm{COSY}}$ are illustrated.}
  \label{fig:example}
\end{figure}

In general, the relative phase $\phi$ between the 
rf signal and the spin precession is time dependent and defined as
\begin{equation}
\phi(t) = 2\pi (t-t_0) (f_\mathrm{rf}  - \nu_s f_\mathrm{cosy} ) + \phi_0 \, ,
\end{equation}
where $t$ is the time in the cycle, $t_0$ is the time the measurement starts and $\phi_0$ is the phase at $t=t_0$. 
As the solenoid is located
at a fixed position in the ring, the time is sampled as $t-t_0 = n/f_\mathrm{cosy}$ 
where $n$ denotes the turn number:
\begin{equation}
\phi(n) = 2\pi n \left( \frac{f_\mathrm{rf}}{f_\mathrm{cosy}} - \nu_s \right) +
\phi_0 \, .
\label{eq:phi_dependence}
\end{equation}
As long as Eq.~(\ref{eq:resonance_condition}) is fulfilled, $\phi(n)$ will remain
constant (modulo $2\pi$), since the term in the parentheses is an integer.
However, even for a small mismatch between the resonance frequency and
$f_\mathrm{rf}$, the value of $\phi$ will continuously change and the polarization build-up will 
be diminished or even canceled. If one accepts the build-up to drop to 80\% at the end of a $\SI{1000}{s}$
cycle (corresponding to $\cos\Delta\phi = 0.8$ and $\Delta\phi \approx \SI{0.65}{rad}$), both frequencies 
have to be adjusted with a precision better than $\Delta f/f = \num{e-10}$. As discussed in
Ref.~\cite{Eversmann:2015jnk}, both the reproducibility and the stability
of the spin tune $\nu_s$ are of the order of \num{e-8} to \num{e-9}.
Figure~\ref{fig:example}(a) provides an example for the variation of $\phi$ over
one cycle without a feedback system. Starting at an arbitrary value the phase is continuously changing
over the whole cycle. With the same initial settings other cycles
show both weaker and stronger time dependencies. Therefore, an active feedback system is essential 
for keeping $\phi$ within the desired interval of $\pm\Delta\phi$. 

\begin{figure}
\includegraphics[width=\columnwidth]{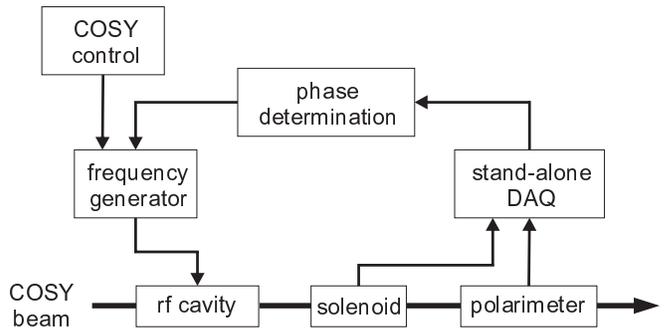}
  \caption{Schematic layout of the feedback system. Based on the 
  measured polarization direction in the polarimeter and the phase of the rf solenoid 
  the control system adds an offset in the range of $\SI{\pm2}{Hz}$ to the nominal value 
  for the revolution frequency.
  The minimum step size is $\SI{3.7}{mHz}$.
  }
  \label{fig:implementation}
\end{figure}

The basic principle of the feedback system is to control the spin 
precession frequency by varying $f_\mathrm{cosy}$ and to match
its phase to the radio frequency of the solenoid.
Figure~\ref{fig:implementation} illustrates schematically the feedback 
control loop. In a first step, the phase of the 
spin precession is measured using a 
polarimeter and a stand-alone data acquisition system as described in 
\cite{Eversmann:2015jnk}. 
Then, the relative 
phase $\phi_\mathrm{meas} 
 = \phi + \phi_\mathrm{det}$ 
is determined, which 
corresponds to the difference between the solenoid phase and the spin precession phase.
Both phases are measured using the same 
reference clock. 
The additional
constant offset $\phi_\mathrm{det}$ originates 
from the relative position of polarimeter
and rf solenoid in the ring and from cable delays. 
The typical measurement
period for each data point is about $\SI{2}{s}$ depending on the available 
intensity of the beam, 
which is continuously extracted onto the target during the cycle.
The purpose of the feedback
is twofold:
First, the phase has to be adjusted to a predefined value $\phi_\mathrm{set}$
to define and to reproduce the experimental conditions. This is done by
running off-resonance for a short period of time (of the order of
$\SI{100}{ms}$, e.g. in Fig.~\ref{fig:example}(b) at $t\approx 90$\,s)
until the desired phase has been reached. It is also used to compensate
larger deviations during the course of the cycle.
Second, the resonance condition, defined in Eq.~(\ref{eq:resonance_condition}) 
has to be fulfilled in order to keep $\phi_\mathrm{meas}$ constant. This
is achieved by introducing a corresponding offset to the revolution frequency $f_\mathrm{cosy}$. 

The necessary frequency change $\Delta f_\mathrm{cosy}$ can be calculated
using the derivative of Eq.~(\ref{eq:phi_dependence}):
\begin{equation}
\frac{\partial\phi}{\partial f_\mathrm{cosy}} = 2\pi n \left( 
-\frac{f_\mathrm{rf}}{f^2_\mathrm{cosy}} - \frac{\partial\nu_s}{\partial 
f_\mathrm{cosy}} \right) \, .
\end{equation}
While the first term originates from the different arrival time at
the solenoid, the second term describes the change of the spin tune:
\begin{equation}
\frac{\Delta \nu_s}{\nu_s} = \frac{\Delta\gamma}{\gamma} = \beta^2 
\frac{\Delta p}{p} = \frac{\beta^2}{\eta} \frac{\Delta
  f_\mathrm{cosy}}{f_\mathrm{cosy}} \, ,
\label{eq:spintune}
\end{equation}
where $\beta$ is the relative particle velocity and $\eta$ the so-called
slip factor, which connects a change in momentum $p$ into a change
in revolution frequency.
Using values given in Tab.~\ref{fig:example} one obtains
\begin{equation}
\Delta\phi \approx 6.93\,\mathrm{\frac{rad}{Hz\,s}} \Delta f_\mathrm{cosy} \Delta t .
\end{equation}
With a minimum step size of $3.7\,\mathrm{mHz}$ offered by the standard
COSY frequency generator it was possible to induce phase adjustments in 
steps of $\Delta\phi \approx \pm 26\,\mathrm{mrad/s}$.

Figure~\ref{fig:example}(b) shows the phase difference $\phi_\mathrm{meas} -
\phi_\mathrm{set}$ versus time while the feedback
system was active. In  Fig.~\ref{fig:example}(c) the changes to the revolution 
frequency are illustrated. The feedback system
started monitoring the phase at $t=\SI{85}{s}$ when the
polarization vector was turned into the horizontal plane. While a first 
correction of the phase drift was applied at the first vertical dashed line, the
pulse at the second dashed line moved the phase to the predefined
value $\phi_{\textrm{set}}$. Once this phase had been reached it was kept stable
over the remaining cycle. The resulting phase distribution has a width of
$\sigma = \SI{0.21}{rad}$, indicated by the shaded band in Fig.~\ref{fig:example}b,
which is well below the allowed variation defined by $\cos\Delta\phi = 0.8$. 

\newcommand{\polDiagram}{
  \begin{tikzpicture}[
      scale=3,
      axis/.style={very thick, ->},
    ]
    \coordinate (p) at (0.75, 0.75);
    \coordinate (pNew) at (0.6, 0.75);
    \coordinate (pHor) at (0.75, 0);
    \coordinate (pHorNew) at (0.6, 0);
    
    \draw[axis] (-0.1,0)  -- (1.1,0) node(xline)[right]
        {$p_H$};
    \draw[axis] (0,-0.1) -- (0,1.1) node(yline)[above] {$p_V$};
    
    \draw[thick, ->] (0,0) to (p) node[right] {$\mathbf{p}$};
    \draw[thick, ->, dashed] (0,0) to (pNew);
    \draw[dashed] (p) to (pHor);
    \draw[dashed] (pNew) to (pHorNew);
    \draw[thick, ->] (p) to (pNew) node[above] {$\Delta p_H$};
    \draw (0.2,0) arc(0:45:0.2) node[below]{$\alpha$};
  \end{tikzpicture}
}

\begin{figure}
  \includegraphics[width=\columnwidth]{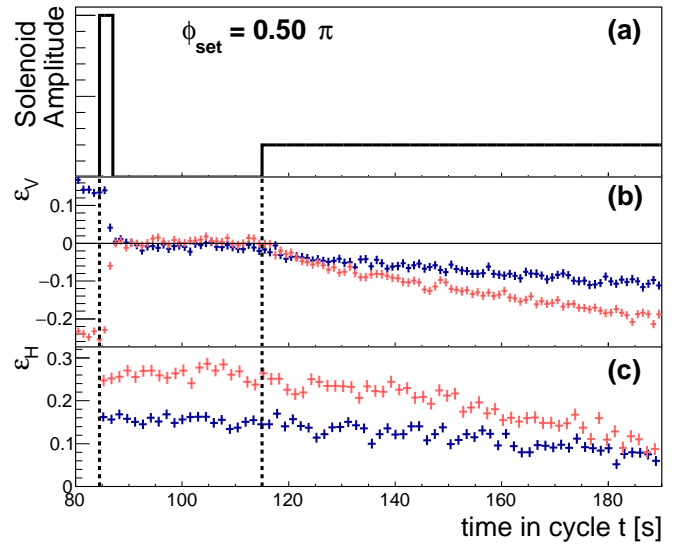}
  \caption{Vertical ($\epsilon_{V}$) and horizontal ($\epsilon_{H}$) asymmetries for positive (blue) and negative (red) initial polarization.
  The polarization is rotated into the horizontal plane by means of the
  solenoid at $t=85\,$s.
  The feedback system is switched on
  and the polarization is slowly rotated back towards the vertical direction
at $t=115\,$s.
  }
  \label{fig:spinFlip}
\end{figure}

Fig.~\ref{fig:example}(b) already proves the functionality of the
feedback system.
In order to confirm that the relative phase 
can really be set to an arbitrary value between $0$ and $2\pi$,
the phase dependent build-up of the vertical polarization
was measured.
The basic principle of the experiment is shown in Fig.~\ref{fig:spinFlip}.

Once the polarization vector had been turned into the horizontal plane
(first vertical dashed line),
the feedback system was switched on, maintaining the relative phase $\phi_\mathrm{meas}$
at a predefined value $\phi_\mathrm{set}=0.5\pi$. At $t=\SI{115}{s}$ (second dashed line)
the solenoid was switched back 
on at a reduced strength causing the polarization to rotate slowly out of the horizontal plane
--- either upwards or downwards depending on the selected value of $\phi = \phi_\mathrm{set} - \phi_\mathrm{det}$.
The field integral $\int{B_\mathrm{eff}\mathrm{d}l}$ of the solenoid was about $\SI{4.6e-4}{Tmm}$.
This corresponds to a maximal polarization rotation of 50~mrad/s.

Figure~\ref{fig:spinFlip}(a) qualitatively shows the amplitude of the solenoid.
Figure~\ref{fig:spinFlip}(b) shows the asymmetry
$\epsilon_V = (N_{\mathrm L} -N_{\mathrm R})/(N_{\mathrm L}+N_{\mathrm R})$
proportional to the polarization in the vertical direction, obtained 
from the counting rates in the left ($N_{\mathrm L}$) and right ($N_{\mathrm R}$) detector segments.
Figure~\ref{fig:spinFlip}(c) shows the asymmetry $\epsilon_H$ 
proportional to the amplitude of the polarization in the horizontal plane.
 $\epsilon_H$  is obtained from the oscillating event rates
in the upper and lower segment of the detector, $N_{\mathrm U}$ and 
$N_{\mathrm D}$ as described in Ref.~\cite{Eversmann:2015jnk}.

\begin{figure}
  \includegraphics[width=\columnwidth]{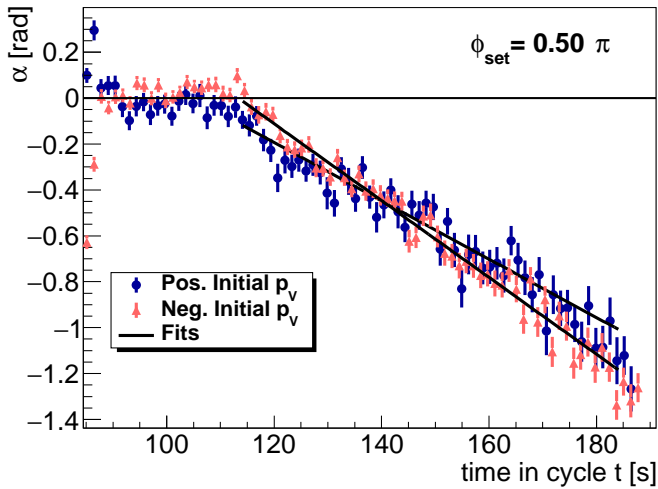}
  \caption{Rotation angle $\alpha$ as function of time since injection for
  positive (blue, circles) and negative (red, triangles) initial 
  polarization.
  \label{fig:fitExample}}
\end{figure}

\begin{figure}
  \includegraphics[width=\columnwidth]{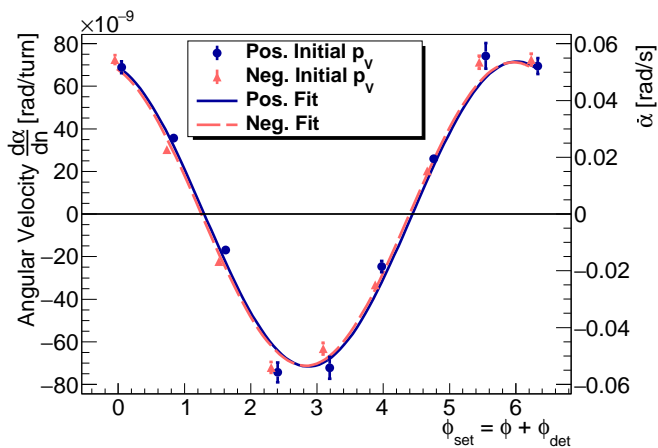}
  \caption{Angular velocity of the solenoid polarization rotation as a function 
  of the nominal phase for data with positive (blue, circles) and negative (red, triangles) 
  initial polarization.
  For better visibility 
  the points were slightly shifted to the left and right horizontally.
}
  \label{fig:sineFit}
\end{figure}

The polarization rotation is analyzed by examining the angle $\alpha=\arctan\left(\epsilon_{V}/\epsilon_{H}\right)$
between the polarization vector and the horizontal plane. $\alpha$
is independent on the initial degree of polarization and
is expected
to either increase or decrease linearly with a slope proportional to $\sin\phi$.
Figure~\ref{fig:fitExample} shows $\alpha$ as function of time in cycle for
$\phi_\mathrm{set} = 0.5\,\pi$.
Both initial polarization states show a similar slope.
Figure~\ref{fig:sineFit} shows the fitted slopes as a function of the nominal phase 
$\phi_\textrm{set} = \phi + \phi_\mathrm{det}$. The rate of build-up proportional to $\sin\phi$ is
another striking proof that the method works.

In conclusion, 
we have shown that the high precision measurement of the spin tune, discussed in \cite{Eversmann:2015jnk}, can
be used to perform a real-time measurement of the relative phase between
the spin precession and an external radio frequency device. This has been used for
the first time to implement 
an active feedback system to stabilize this phase at a level of $\SI{0.21}{rad}$
and to reproduce the phase dependence of the polarization rotation in an rf solenoid.
As the sensitivity to phase variations is the same for the rf Wien filter, 
this is an essential prerequisite towards the first measurement of
an EDM in a storage ring. 
In addition, such a feedback system can also be used in the frozen-spin
method~\cite{Anastassopoulos:2015ura},  where the polarization has to be kept aligned to the momentum vector in the
horizontal plane.

\begin{acknowledgments}
The authors wish to thank the staff of COSY for providing excellent working 
conditions and for their support concerning the technical aspects
of this experiment. This work has been financially supported by  
Forschungszentrum J\"ulich GmbH, Germany, via COSY FFE, 
by an ERC Advanced-Grant (srEDM \# 694390) of the European Union,
the European Union Seventh Framework Programme (FP7/2007-2013) under Grant 
Agreement No. 283286, 
the Shota Rustaveli National Science Foundation of the Republic of 
Georgia, by a Grant from the
Russian Science Foundation (grant number RNF-16-12-10151), and
by IBS-R017-D1-2017-a00.

\end{acknowledgments}

\bibliographystyle{apsrev4-1}
\bibliography{literature_edm}

\begin{thebibliography}{10}%
\makeatletter
\providecommand \@ifxundefined [1]{%
 \@ifx{#1\undefined}
}%
\providecommand \@ifnum [1]{%
 \ifnum #1\expandafter \@firstoftwo
 \else \expandafter \@secondoftwo
 \fi
}%
\providecommand \@ifx [1]{%
 \ifx #1\expandafter \@firstoftwo
 \else \expandafter \@secondoftwo
 \fi
}%
\providecommand \natexlab [1]{#1}%
\providecommand \enquote  [1]{``#1''}%
\providecommand \bibnamefont  [1]{#1}%
\providecommand \bibfnamefont [1]{#1}%
\providecommand \citenamefont [1]{#1}%
\providecommand \href@noop [0]{\@secondoftwo}%
\providecommand \href [0]{\begingroup \@sanitize@url \@href}%
\providecommand \@href[1]{\@@startlink{#1}\@@href}%
\providecommand \@@href[1]{\endgroup#1\@@endlink}%
\providecommand \@sanitize@url [0]{\catcode `\\12\catcode `\$12\catcode
  `\&12\catcode `\#12\catcode `\^12\catcode `\_12\catcode `\%12\relax}%
\providecommand \@@startlink[1]{}%
\providecommand \@@endlink[0]{}%
\providecommand \url  [0]{\begingroup\@sanitize@url \@url }%
\providecommand \@url [1]{\endgroup\@href {#1}{\urlprefix }}%
\providecommand \urlprefix  [0]{URL }%
\providecommand \Eprint [0]{\href }%
\providecommand \doibase [0]{http://dx.doi.org/}%
\providecommand \selectlanguage [0]{\@gobble}%
\providecommand \bibinfo  [0]{\@secondoftwo}%
\providecommand \bibfield  [0]{\@secondoftwo}%
\providecommand \translation [1]{[#1]}%
\providecommand \BibitemOpen [0]{}%
\providecommand \bibitemStop [0]{}%
\providecommand \bibitemNoStop [0]{.\EOS\space}%
\providecommand \EOS [0]{\spacefactor3000\relax}%
\providecommand \BibitemShut  [1]{\csname bibitem#1\endcsname}%
\let\auto@bib@innerbib\@empty
\bibitem [{jed()}]{jedi}%
  \BibitemOpen
  \href@noop {} {}\bibinfo {note} {{JEDI Collaboration,
  \url{http://collaborations.fz-juelich.de/ikp/jedi/index.shtml}}}\BibitemShut
  {NoStop}%
\bibitem [{\citenamefont {Engel}\ \emph {et~al.}(2013)\citenamefont {Engel},
  \citenamefont {Ramsey-Musolf},\ and\ \citenamefont {van
  Kolck}}]{Engel201321}%
  \BibitemOpen
  \bibfield  {author} {\bibinfo {author} {\bibfnamefont {J.}~\bibnamefont
  {Engel}}, \bibinfo {author} {\bibfnamefont {M.~J.}\ \bibnamefont
  {Ramsey-Musolf}}, \ and\ \bibinfo {author} {\bibfnamefont {U.}~\bibnamefont
  {van Kolck}},\ }\href {\doibase http://dx.doi.org/10.1016/j.ppnp.2013.03.003}
  {\bibfield  {journal} {\bibinfo  {journal} {Progress in Particle and Nuclear
  Physics}\ }\textbf {\bibinfo {volume} {71}},\ \bibinfo {pages} {21 }
  (\bibinfo {year} {2013})}\BibitemShut {NoStop}%
\bibitem [{\citenamefont {Canetti}\ \emph {et~al.}(2012)\citenamefont
  {Canetti}, \citenamefont {Drewes},\ and\ \citenamefont
  {Shaposhnikov}}]{Canetti:2012zc}%
  \BibitemOpen
  \bibfield  {author} {\bibinfo {author} {\bibfnamefont {L.}~\bibnamefont
  {Canetti}}, \bibinfo {author} {\bibfnamefont {M.}~\bibnamefont {Drewes}}, \
  and\ \bibinfo {author} {\bibfnamefont {M.}~\bibnamefont {Shaposhnikov}},\
  }\href {\doibase 10.1088/1367-2630/14/9/095012} {\bibfield  {journal}
  {\bibinfo  {journal} {New J. Phys.}\ }\textbf {\bibinfo {volume} {14}},\
  \bibinfo {pages} {095012} (\bibinfo {year} {2012})},\ \Eprint
  {http://arxiv.org/abs/1204.4186} {arXiv:1204.4186 [hep-ph]} \BibitemShut
  {NoStop}%
\bibitem [{\citenamefont {Guidoboni}\ \emph {et~al.}(2016)\citenamefont
  {Guidoboni} \emph {et~al.}}]{Guidoboni:2016bdn}%
  \BibitemOpen
  \bibfield  {author} {\bibinfo {author} {\bibfnamefont {G.}~\bibnamefont
  {Guidoboni}} \emph {et~al.} (\bibinfo {collaboration} {JEDI}),\ }\href
  {\doibase 10.1103/PhysRevLett.117.054801} {\bibfield  {journal} {\bibinfo
  {journal} {Phys. Rev. Lett.}\ }\textbf {\bibinfo {volume} {117}},\ \bibinfo
  {pages} {054801} (\bibinfo {year} {2016})}\BibitemShut {NoStop}%
\bibitem [{\citenamefont {Eversmann}\ \emph {et~al.}(2015)\citenamefont
  {Eversmann} \emph {et~al.}}]{Eversmann:2015jnk}%
  \BibitemOpen
  \bibfield  {author} {\bibinfo {author} {\bibfnamefont {D.}~\bibnamefont
  {Eversmann}} \emph {et~al.} (\bibinfo {collaboration} {JEDI}),\ }\href
  {\doibase 10.1103/PhysRevLett.115.094801} {\bibfield  {journal} {\bibinfo
  {journal} {Phys. Rev. Lett.}\ }\textbf {\bibinfo {volume} {115}},\ \bibinfo
  {pages} {094801} (\bibinfo {year} {2015})},\ \Eprint
  {http://arxiv.org/abs/1504.00635} {arXiv:1504.00635 [physics.acc-ph]}
  \BibitemShut {NoStop}%
\bibitem [{\citenamefont {Morse}\ \emph {et~al.}(2013)\citenamefont {Morse},
  \citenamefont {Orlov},\ and\ \citenamefont
  {Semertzidis}}]{PhysRevSTAB.16.114001}%
  \BibitemOpen
  \bibfield  {author} {\bibinfo {author} {\bibfnamefont {W.~M.}\ \bibnamefont
  {Morse}}, \bibinfo {author} {\bibfnamefont {Y.~F.}\ \bibnamefont {Orlov}}, \
  and\ \bibinfo {author} {\bibfnamefont {Y.~K.}\ \bibnamefont {Semertzidis}},\
  }\href {\doibase 10.1103/PhysRevSTAB.16.114001} {\bibfield  {journal}
  {\bibinfo  {journal} {Phys. Rev. ST Accel. Beams}\ }\textbf {\bibinfo
  {volume} {16}},\ \bibinfo {pages} {114001} (\bibinfo {year}
  {2013})}\BibitemShut {NoStop}%
\bibitem [{\citenamefont {Slim}\ \emph {et~al.}(2016)\citenamefont {Slim} \emph
  {et~al.}}]{Slim:2016pim}%
  \BibitemOpen
  \bibfield  {author} {\bibinfo {author} {\bibfnamefont {J.}~\bibnamefont
  {Slim}} \emph {et~al.},\ }\href {\doibase 10.1016/j.nima.2016.05.012}
  {\bibfield  {journal} {\bibinfo  {journal} {Nucl. Instrum. Meth.}\ }\textbf
  {\bibinfo {volume} {A828}},\ \bibinfo {pages} {116} (\bibinfo {year}
  {2016})},\ \Eprint {http://arxiv.org/abs/1603.01567} {arXiv:1603.01567
  [physics.ins-det]} \BibitemShut {NoStop}%
\bibitem [{\citenamefont {Maier}(1997)}]{Maier:1997zj}%
  \BibitemOpen
  \bibfield  {author} {\bibinfo {author} {\bibfnamefont {R.}~\bibnamefont
  {Maier}},\ }\href {\doibase 10.1016/S0168-9002(97)00324-0} {\bibfield
  {journal} {\bibinfo  {journal} {Nucl. Instrum. Meth.}\ }\textbf {\bibinfo
  {volume} {A390}},\ \bibinfo {pages} {1} (\bibinfo {year} {1997})}\BibitemShut
  {NoStop}%
\bibitem [{\citenamefont {Bagdasarian}\ \emph {et~al.}(2014)\citenamefont
  {Bagdasarian} \emph {et~al.}}]{PhysRevSTAB.17.052803}%
  \BibitemOpen
  \bibfield  {author} {\bibinfo {author} {\bibfnamefont {Z.}~\bibnamefont
  {Bagdasarian}} \emph {et~al.},\ }\href {\doibase
  10.1103/PhysRevSTAB.17.052803} {\bibfield  {journal} {\bibinfo  {journal}
  {Phys. Rev. ST Accel. Beams}\ }\textbf {\bibinfo {volume} {17}},\ \bibinfo
  {pages} {052803} (\bibinfo {year} {2014})}\BibitemShut {NoStop}%
\bibitem [{\citenamefont {Anastassopoulos}\ \emph {et~al.}(2016)\citenamefont
  {Anastassopoulos} \emph {et~al.}}]{Anastassopoulos:2015ura}%
  \BibitemOpen
  \bibfield  {author} {\bibinfo {author} {\bibfnamefont {V.}~\bibnamefont
  {Anastassopoulos}} \emph {et~al.},\ }\href@noop {} {\bibfield  {journal}
  {\bibinfo  {journal} {Rev. Sci. Instrum.}\ }\textbf {\bibinfo {volume}
  {87}},\ \bibinfo {pages} {115116} (\bibinfo {year} {2016})},\ \Eprint
  {http://arxiv.org/abs/1502.04317} {arXiv:1502.04317 [physics.acc-ph]}
  \BibitemShut {NoStop}%
\end{thebibliography}%
\end{document}